\documentclass[figures]{epl}
\usepackage{definitions}
\usepackage{localdefs}
\usepackage{amsmath}
\usepackage{amssymb}
\usepackage{graphics}
\usepackage{amsbsy}
\usepackage{bm}
\usepackage{cite}

\title{Phase equilibria in stratified thin liquid films stabilized by
colloidal particles}

\author{J.~B{\l}awzdziewicz\inst{1} and  E.~Wajnryb\inst{1,2}}
\institute{\inst{1} Department of Mechanical Engineering, Yale University,
P.O. Box 20-8286, New Haven, CT 06520, U.S.A.
\\
\inst{2} On leave from IPPT Warsaw, Poland.
}

\pacs{68.15.+e}{Liquid thin films}
\pacs{68.18.Jk}{Phase transitions}
\pacs{82.70.Dd}{Colloids}

\begin{document}

\maketitle

\begin{abstract}

Phase equilibria between regions of different thickness in thin liquid
films stabilized by colloidal particles are investigated using a
quasi-two-dimensional thermodynamic formalism.  
Appropriate equilibrium conditions for the film tension, normal
pressure, and chemical potential of the particles in the film are
formulated, and it is shown that the relaxation of these parameters
occurs consecutively on three distinct time scales.  Film
stratification is described quantitatively for a hard-sphere
suspension using a Monte-Carlo method to evaluate thermodynamic
equations of state.  Coexisting phases are determined for systems in
constrained- and full-equilibrium states that correspond to different
stages of film relaxation.

\end{abstract}

Drainage of thin liquid films often involves spontaneous formation of
coexisting regions of different, but uniform, thickness. For example,
circular black-film spots can be seen in soap bubbles prior to
breakup.  Such a stratification phenomenon is particularly interesting
in films stabilized by colloidal particles, micelles, or
polyelectrolytes, where the thinning occurs through a series of
stepwise transitions between film states characterized by the
thickness commensurate with the size of the stabilizing particles.  In
experiments with horizontal films
\cite{%
Nikolov-Wasan:1992,%
Sethumadhavan-Nikolov-Wasan:2001%
},
such transitions occur by rapid formation of circular regions of a
smaller thickness, followed by a much slower expansion.  In
investigations of vertical films, up to seven coexisting parallel
stripes of different thickness have been observed
\cite{Basheva-Danov-Kralchevsky:1997}. 

Film stratification has recently been intensively studied
\cite{%
Nikolov-Wasan-Kralchevsky-Ivanov:1992,%
Nikolov-Wasan:1992,%
Basheva-Danov-Kralchevsky:1997,%
Bergeron-Radke:1992,%
Theodoly-Tan-Ober-Williams-Bergeron:2001,%
Sethumadhavan-Nikolov-Wasan:2001,
Stubenrauch-von_Klitzing:2003,%
Bergeron:1999%
}
because of the relevance of the problem for understanding structural
colloidal forces.  In particle-stabilized thin liquid films, the
structural forces depend on the film thickness in an oscillatory
manner due to particle layering \cite{Kralchevsky-Danov-Ivanov:1996}.
Equilibrium values of the film thickness result from the balance
between the normal pressure in the film (which includes the
oscillatory structural force) and the outside pressure
\cite{%
Theodoly-Tan-Ober-Williams-Bergeron:2001,%
Trokhymchuk-Henderson-Nikolov-Wasan:2001,%
Stubenrauch-von_Klitzing:2003%
}.
We find, however, that the normal-stress balance used alone in
descriptions of the film stratification phenomenon is
insufficient---the mechanical equilibrium between regions (phases) of
different thickness requires also the lateral force balance.  Since
the lateral balance has not been included in the available theories,
understanding of an essential aspect of the problem is still lacking.
One of our goals is to provide a proper thermodynamic framework for
analyzing the phase behavior of particle stabilized thin liquid films.

An analysis of the stepwise-thinning phenomenon requires equilibrium
as well as nonequilibrium considerations.  Experimental observations
indicate that the thinning process occurs on several distinct time
scales: a thinner spot nucleates in a fraction of a second but the
time scale for the subsequent expansion of the spot is much slower.
This separation of the time scales suggests that the system at
different evolution stages is, approximately, in {\it constrained
equilibrium\/}.  In such states the film phases of uniform thickness
are in equilibrium.  However, only some of the equilibrium conditions
between these phases are satisfied---the others are not, due to
dynamical constraints.  An important goal of our study is to identify
relevant time scales in the stepwise-thinning phenomenon, and to
discern the corresponding relaxation processes.  Our theoretical
considerations are supplemented with results of Monte-Carlo
simulations of a model system to illustrate possible full and partial
equilibrium states of a stratified particle-stabilized film at
different evolution stages.

Our analysis of the film stratification phenomenon relies on
representing the film as an effective two-dimensional medium
\cite{%
Kralchevsky-Danov-Ivanov:1996,%
Ikeda-Krustev-Muller:2004%
}. 
Accordingly, all details of the transverse structure of the film are
averaged out (as in the thermodynamic description of an interface
proposed by Gibbs). The film thickness, however, is retained as a
parameter of state, because it can be controlled by varying the
difference between the external pressure $\externalP$ and the pressure
of the bulk fluid in contact with the fluid in the film.  We note that
the usual formulation of thin-film thermodynamics
\cite{Kralchevsky-Danov-Ivanov:1996} is insufficient for our problem
because of the underlying assumptions that the pressure in the film is
isotropic and the interaction between the film interfaces is described
by a mean-field normal force.  In contrast, the stabilizing colloidal
particles contribute not only to the normal but also to the tangential
stress balance in the film, and the corresponding stress tensor is
non-isotropic due to confinement effects.  These features have to be
taken into account in the proper formulation of the theory.

\enlargethispage{5pt}
A typical thickness of a particle-stabilized liquid film is in the
micrometer range, consistent with the size of colloidal particles.  On
this scale, the film interfaces can be treated as surfaces with zero
thickness and interfacial tension $\sigma$.  Assuming this geometry, a
suitable effective two-dimensional thermodynamic description can be
obtained from the expression for the macroscopic infinitesimal work
\begin{equation}
\label{work}
\dd W=-A\normalP\diff h+(2\sigma-h\lateralPaver)\diff A
\end{equation}
associated with changes of the film thickness $h$ and area $A$.  In
the above relation, $\normalP$ and $\lateralP$ denote the normal and
lateral components of the pressure tensor in the film and the overbar
indicates the average value across the film.  If a direct interaction
force ({\it e.g.\/}, the van der Waals attraction or electrostatic
repulsion) acts between the film interfaces, the normal pressure
$\normalP$ in eq.~\refeq{work} should be replaced with $\normalP+f_n$,
where $f_n$ is the normal force per unit area, and $f_n>0$ corresponds
to attraction.

\enlargethispage{12pt} In eq.\ \refeq{work}, the work associated with the
area change consists of the interfacial-tension and pressure
contributions.  To obtain an effective two-dimensional thermodynamic
description of the film, the lateral interfacial-tension forces and
the pressure are combined into a single conjugate thermodynamic
variable. However, first an appropriate set of independent
thermodynamic parameters has to be chosen.  The pair of independent
variables $h$ and $A$ used in Eq.\ \refeq{work} is unsuitable because
$h$ is not an extensive quantity. By a simple change of variables, a
more appropriate form is obtained
\begin{equation}
\label{work in terms of film tension}
\dd W=-\normalP\diff V+\tension\diff A,
\end{equation}
where $V=Ah$ is the film volume, and, by analogy with the definition
of the surface tension, 
\begin{equation}
\label{film tension in terms of stress components}
\tension=h(\normalP-\lateralPaver)+2\sigma
\end{equation} 
is the tension of the film \cite{Ikeda-Krustev-Muller:2004}.  

The fundamental relation for the film free energy $F$ can be derived
in the usual way from the expression for mechanical work \refeq{work
in terms of film tension}.  Treating the suspension in the film as a
two-component fluid (with the colloidal particles regarded as
macromolecules) we get
\begin{equation}
\label{fundamental relation for film free energy}
\diff F=-S\diff T-\normalP\diff V+\tension\diff A,
       +\muC\diff \NC+\muF\diff \NF,
\end{equation}
where $\NC$ and $\NF$ denote the number of colloidal particles and
solvent molecules in the system, and $\muC$ and $\muF$ are the
corresponding chemical potentials.  The film-tension representation
\refeq{fundamental relation for film free energy} of the film free
energy is more suitable for discussion of phase equilibria in
stratified films than the more common film surface tension
representation where the film thickness $h$ is retained as an
independent thermodynamic parameter. Since all independent parameters
in our representation $V,A,\NC$ and $\NF$ are extensive, the conjugate
intensive functions of state $\normalP,\tension,\muC$, and $\muF$ in
coexisting phases must be equal in unconstrained equilibrium (which
follows from standard entropy-maximization arguments
\cite{Callen:1985}).  Accordingly, apart from the trivial
thermal-equilibrium condition $T^{(1)}=T^{(2)}=T^\external$, there are
the transverse and lateral mechanical equilibrium conditions
\refstepcounter{equation}
\label{vertical and lateral mechanical equilibrium conditions}
\newcommand{\mechEQ}{\theequation}
\begin{equation}
\tag{\arabic{equation}a,b}
\normalP^{(1)}=\normalP^{(2)}=\externalP,\qquad
\tension^{(1)}=\tension^{(2)},
\end{equation}
and the chemical equilibrium conditions
\refstepcounter{equation}
\label{chemical equilibrium conditions}
\newcommand{\chemEQ}{\theequation}
\begin{equation}
\tag{\arabic{equation}a,b}
\muC^{(1)}=\muC^{(2)}, \qquad\muF^{(1)}=\muF^{(2)},
\end{equation}
where $T^\external$ and $\externalP$ are the external temperature and
pressure.  

A typical size of colloidal particles used in experiments with
stratified thin films (or the range of the effective screened
Coulombic potential if the particles are charged) is
$\gtrsim0.1\,\mu$m.  Under these conditions the solvent can be treated
as a structureless incompressible continuous medium that does not
affect thermodynamic properties of the system. In our further
considerations the solvent variables $\muF$ and $\NF$ are thus
neglected (which is a usual approximation), and the equilibrium
conditions \refeq{vertical and lateral mechanical equilibrium
conditions} are rephrased in terms of the excess quantities associated
with the particle presence, \refstepcounter{equation}
\label{equilibrium conditions for excess quantities}
\begin{equation}
   \normalPCi{1}=\normalPCi{2},
    \qquad\tensionCi{1}=\tensionCi{2},
\tag{\theequation a,b}
\end{equation}
where the index $\colloid$ indicates the particle contributions.

%
As demonstrated in numerous experiments, formation of a stepwise
structure in particle-stabilized films is much faster than the
subsequent evolution of the coexisting phases
\cite{%
Nikolov-Wasan:1992,%
Basheva-Danov-Kralchevsky:1997,%
Sethumadhavan-Nikolov-Wasan:2001%
}.
Scaling arguments, outlined below, suggest that the relaxation of a
stratified film is characterized by three distinct time scales
$\tauL\ll\tauT\ll\tauC$ for attaining the equilibrium conditions
\refeqa{equilibrium conditions for excess quantities}{b},
\refeqa{equilibrium conditions for excess quantities}{a}, and
\refeqa{chemical equilibrium conditions}{a}, respectively.  The
corresponding nonequilibrium thermodynamic processes entail changes in
the areas of individual phases, their volumes, and the volume fraction
of colloidal particles.

Estimates of the characteristic relaxation times can be obtained using
energy-dissipation arguments.  To evaluate the time scale $\tauL$ we
note that the area change involves energy dissipation in the whole
volume of the film $V$ and is characterized by the velocity variation
on the lengthscale $l$ set by the lateral film dimension.  By
comparing the energy-dissipation rate $V\eta(u/l)^2$ to the power
$ulh\nC kT$ associated with the thermal stresses $\nC kT$ produced by
the particles, we find $\tauL\sim\eta/\nC kT$, where $\eta$ is the
fluid viscosity, $u\sim\l/\tauL$ is the characteristic film velocity,
$\nC=\NC/V$, and $kT$ is the thermal energy.  The change in the volume
of the coexisting phases requires suspension flow through the contact
region where the film thickness changes rapidly.  Accordingly, the
magnitude of the velocity gradient is $O(u/h)$ and the dissipation
occurs in the volume $(h/l)V$, which yields the estimate $\tauT\sim
(l/h)\tauL$.  A change of the particle volume fraction
$\volumeFractionC$ involves particle motion with respect to the fluid
in the whole volume of the film.  Assuming $d\sim h$ (where $d$ is the
particle diameter) and $\volumeFractionC=O(1)$ we find that $\tauC\sim
(l/h)\tauT$.  For an aqueous suspension of particles with the diameter
$d=0.1$ $\mu$m in a film with the characteristic lateral dimension
$l=10$ $\mu$m we find $\tauL\sim10^{-3}$ s, $\tauT\sim10^{-1}$ s, and
$\tauC\sim10$~s, which is consistent with experimental observations
\cite{Sethumadhavan-Nikolov-Wasan:2001}.  We note that our scaling
analysis applies to films with surfactant-free interfaces (as in the
experiments reported in \cite{Sethumadhavan-Nikolov-Wasan:2001}). We
expect that a surfactant adsorbed on film interfaces may increase the
time scale $\tauT$ by a factor $l/h$, due to the modification of
hydrodynamic boundary conditions.  Also, the timescale $\tauC$ may be
decreased by a factor $d/h$ for charged particles with $d\ll h$.

Assuming the separation of time scales $\tauL\ll\tauT\ll\tauC$ we
anticipate the following scenario for the stepwise-thinning of the
film.  At an early stage of phase separation, the relaxation occurs on
the time scale $\tauL$ primarily through a change in the areas of the
coexisting phases at constant $\volumeFractionC$.  The process is
driven by an imbalance of lateral forces, and after it is completed
the film is in a constrained equilibrium state where only the
lateral-force equilibrium condition \refeqa{equilibrium conditions for
excess quantities}{b} between the coexisting phases is satisfied.  (An
analogy is the coexistence of water and air where the pressure
difference between the two phases relaxes very fast, but equilibration
of the temperature and chemical-potential differences is much slower.)
The next stage of the film evolution is driven by the difference
$\Delta\normalPC$ of the colloidal contribution to the normal pressure
in the coexisting regions.  As required by the incompressibility of
the suspension in the film, this pressure difference is compensated by
the dynamic pressure drop due to the suspension flow through the
contact region.  After the pressure difference $\Delta\normalPC$ has
relaxed on the time scale $\tauT$ the chemical potential of colloidal
particles relaxes on the time-scale $\tauC$ due to the particle
diffusion in the film.  Only this last process entails a change of
particle volume fraction in the coexisting regions.

\begin{figure}
%
\twoimages[width=74mm]{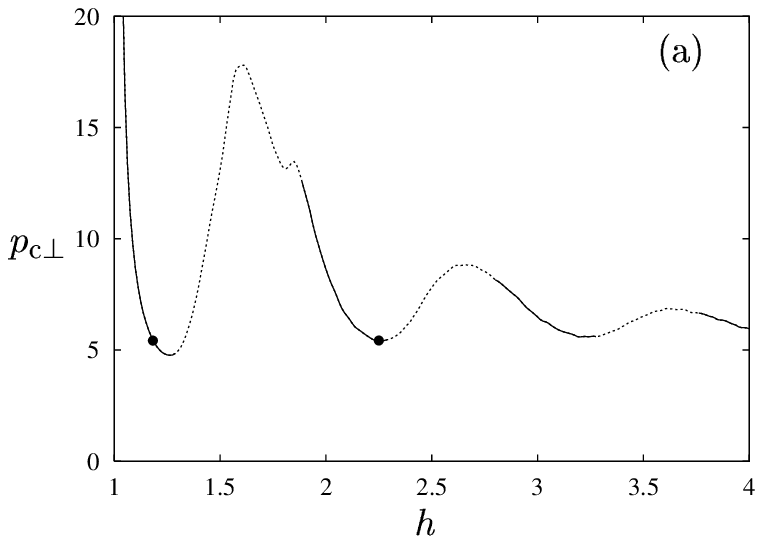}{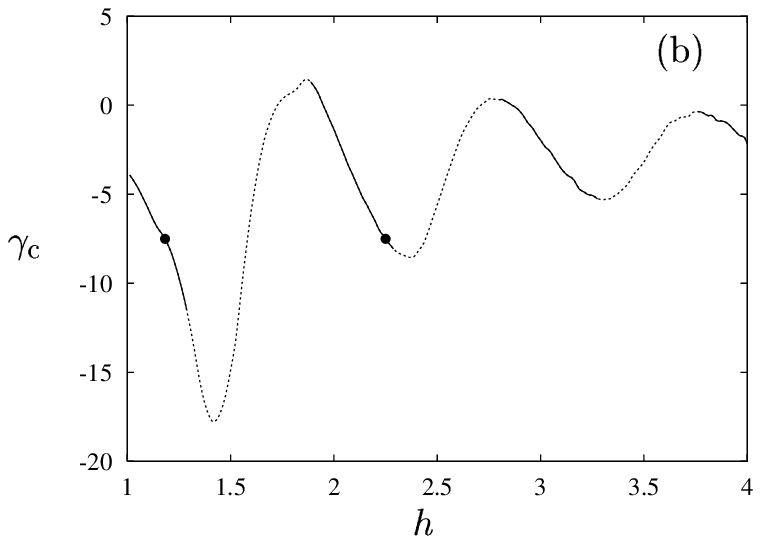}
\caption{
Particle contribution to normal pressure $\normalPC$ and film tension
$\tensionC$ versus film thickness $h$ for particle volume fraction
$\volumeFractionC=0.4$.  Dotted lines correspond to states for which
the stability condition (\protect\ref{stability condition}) is
violated.  Circles indicate a pair of stable phases in mechanical
equilibrium (\protect\ref{equilibrium conditions for excess
quantities}).
}
\label{pressure and gamma versus h}
\end{figure}

Our theory can be used to obtain quantitative predictions for the
trajectory in the thermodynamic parameter space for a stratified film
during the equilibration process.  To this end the general
thermodynamic relations have to be supplemented with equilibrium
equations of state.  We present numerical results for a film
stabilized by a monodisperse suspension of hard spheres.  The particle
contribution to average stress tensor $\averPCtensor$ in homogeneous
phases is evaluated from the contact value of the equilibrium
two-particle density $n_2^{\rm eq}(\br_1,\br_2)$ in the confined
suspension using the relation
\begin{equation}
\label{particle contribution to stress}
\frac{\averPCtensor}{kT}=\nC'
   +\half d^3\,\left\langle\int \hat \br\hat\br \,
     n_2^{\rm eq}(\br_1,\br_1-d\hat \br)\,\diff^2\hat r\right\rangle_{V'}.
\end{equation}
Here $\hat\br=\br/r$, the integration is over the contact
configurations, $\langle\cdots\rangle_{V'}$ denotes the volume average
over the region $V'=A(h-d)$ accessible to particle centers, and
$\nC'=\NC/V'$ is the corresponding particle number
density\footnote{The accessible volume $V'$ rather than the total
volume $V$ appears in eq.~\refeq{particle contribution to stress}
because $\averPCtensor$ is the average of the local stress tensor
$\PCtensor(\br)$ that vanishes for $\br\not\in V'$.  Consistently, we
use the relation $\tensionC=h'(\lateralPCaver-\normalPCaver)$ for the
particle contribution to the film tension.}.  Equation \refeq{particle
contribution to stress} generalizes a well-known expression for the
pressure in a bulk hard-sphere system in terms of the contact value of
the radial distribution.  It can be derived either from the
collisional contribution to the momentum flux or by passing to the
hard-sphere limit of the Kirkwood-Buff expression
\cite{Kirkwood-Buff:1949} for the stress tensor in an inhomogeneous
fluid.  The equilibrium average involved in \refeq{particle
contribution to stress} was evaluated via computer simulations using
the standard Metropolis algorithm to generate the canonical ensemble
for a system of several hundred spheres confined in a planar film with
periodic boundary conditions in the lateral directions.

\begin{figure}
\onefigure[width=76mm]{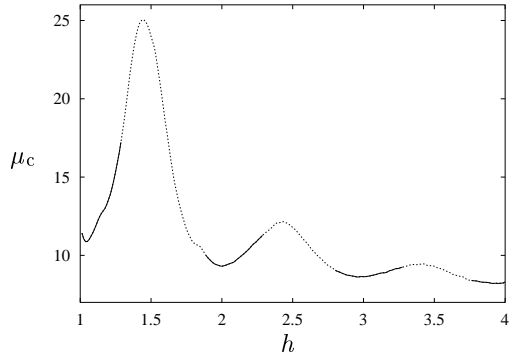}
\caption{
Chemical potential of the particles $\muC$ versus film thickness $h$
for $\volumeFractionC=0.4$.  The normalization corresponds to the
asymptotic behavior $\muC=-\ln v$ for $\volumeFractionC\ll1$. Dotted
line as in fig.~\ref{pressure and gamma versus h}.
}
\label{chemical potential versus h}
\end{figure}

%
A typical dependence of the normal pressure $\normalPC$ and film
tension $\tensionC$ on the film thickness $h$ at a constant volume
fraction $\volumeFractionC$ is shown in fig.~\ref{pressure and gamma
versus h}. By integrating the Gibbs--Duhem relation at fixed $T$
\begin{equation}
\label{Gibbs--Duhem relation}
\diff\muC=v\diff\normalPC-a\diff\tensionC
\end{equation}
(where $v=V/\NC$ and $a=A/\NC$), we also determined the chemical
potential $\muC$.  The plot of $\muC$ versus $h$ at constant
$\volumeFractionC$ is presented in fig.~\ref{chemical potential versus
h}.  The film thickness $h$ is normalized here by the particle
diameter $d$, pressure $\normalPC$ by $kT/d^3$, film tension
$\tensionC$ by $kT/d^2$, and chemical potential $\muC$ by $kT$.  As
expected, the intensive parameters $\normalPC$, $\tensionC$, and
$\muC$ oscillate as functions of the film thickness $h$.  Simulation
results at different volume fractions (not shown) indicate that the
oscillations are more pronounced at higher $\volumeFractionC$.  We
note that the thermodynamic stability of the system requires that
\begin{equation}
\label{stability condition}
(\partial\normalPC/\partial h)_{T\muC}<0,
\end{equation}
which is equivalent to $(\partial\tensionC/\partial h)_{T\muC}<0$.
The domains where \refeq{stability condition} is violated are
indicated in the figures by dotted lines.

%
We now return to our discussion of phase equilibria at different
stages of film evolution.  As predicted by our scaling analysis, the
first two stages occur at a constant volume fraction
$\volumeFractionC^{(1)}=\volumeFractionC^{(2)}$.  At the end of the
first stage, the coexisting phases fulfill the lateral
mechanical-equilibrium condition \refeqa{equilibrium conditions for
excess quantities}{b}; accordingly, the film thickness in these phases
satisfies equation $\tensionC(h)=\tension_0$, where the film tension
is evaluated at constant $\volumeFractionC$, as in fig.\ \ref{pressure
and gamma versus h}b.  The value of the constant $\tension_0$ is not
determined by the equilibrium condition \refeqa{equilibrium conditions
for excess quantities}{b} alone, but it also depends on the initial
volumes of the coexisting regions.  The oscillatory character of the
curve $\tensionC=\tensionC(h)$ in fig.\ \ref{pressure and gamma versus
h}b indicates that several different combinations of coexisting phases
are possible.

\begin{figure}
\twofigures[width=74mm]{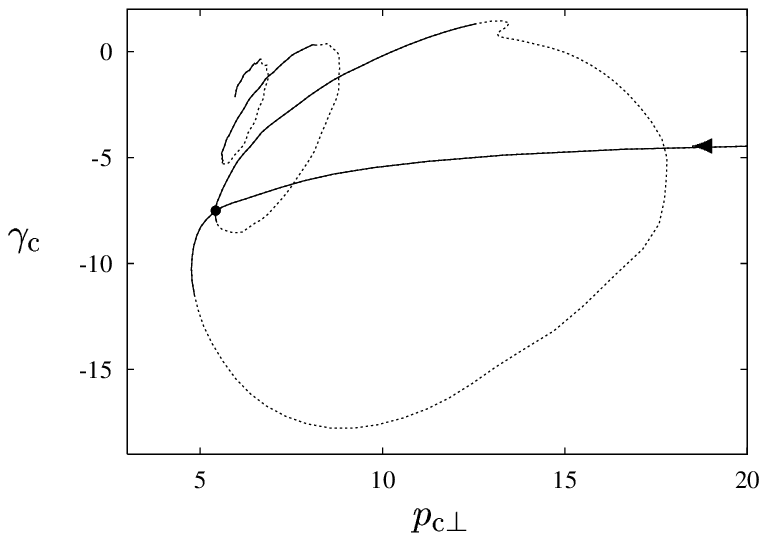}{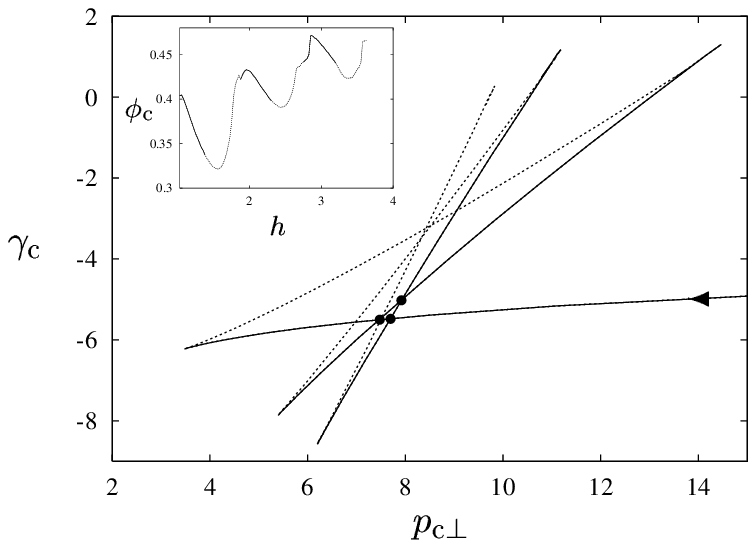}
\caption{
Parametric plot of the equation of state in the
$(\normalPC,\tensionC)$ plane at constant particle volume fraction
$\volumeFractionC=0.4$.  The curve is parametrized by film thickness
$h$; direction of the increasing $h$ is indicated by the arrow.  Dotted
line as in fig.~\protect\ref{pressure and gamma versus h}.  Circle
indicates the stable coexistence state.
}
\label{volume fraction loops}
\caption{
Same as fig.~\protect\ref{volume fraction loops}, except that the
results are for the constant chemical potential $\muC=11$.  Inset
shows the variation of the particle volume fraction with the film
thickness $h$ along the curve.  Circles indicate stable coexistence
states; film thickness in the coexisting phases is
$(h_1,h_2)=(1.1,3.0)$, $(1.1,2.1)$, and $(2.0,3.0)$.
}
\label{chemical potential loops}
\end{figure}

At the next relaxation stage, the complete mechanical equilibrium is
attained.  Since two equilibrium conditions \refeq{equilibrium
conditions for excess quantities} are now satisfied, the coexisting
phases correspond to isolated points in the space of states for given
$\volumeFractionC$.  The coexisting phases can be determined from the
equations of state $\normalPC=\normalPC(h)$ and
$\tensionC=\tensionC(h)$ by combining them into a single parametric
plot, as shown in fig.\ \ref{volume fraction loops} for
$\volumeFractionC=0.4$.  The coexisting phases correspond to the line
intersections.  In our example only one intersection (indicated by the
circle) represents a stable system.  The two coexisting phases are
also marked in fig.\ \ref{pressure and gamma versus h}.  Additional
numerical calculations show that at higher volume fractions several
distinct stable states with coexisting phases can be identified, and
for lower $\volumeFractionC$ there are no such states.  
We emphasize that not only do the intersection points in fig.\
\ref{volume fraction loops} have a physical meaning but also the curve
itself does---its different portions represent the trajectory of the
system in the space of states when the normal-pressure difference
relaxes.  

A similar parametric plot can be used to determine the possible
coexisting phases in full thermodynamic equilibrium.  However, the
chemical potential $\muC$ is now held constant, instead of
$\volumeFractionC$, due to the equilibrium condition \refeqa{chemical
equilibrium conditions}{a}.  An example of such a plot is presented in
fig.~\ref{chemical potential loops}; the inset shows the corresponding
variation of the volume fraction $\volumeFractionC$ with the film
thickness $h$.  The shape of the curve $\muC=\mbox{const}$ in the
$(\normalPC,\tensionC)$ plane is entirely different from the shape of
the line $\volumeFractionC=\mbox{const}$ in fig.~\ref{volume fraction
loops}.  The distinct qualitative features of the curve plotted in
fig.\ \ref{chemical potential loops} follow from Frumkin
\cite{Frumkin:1938} equation
$(\partial\tensionC/\partial\normalPC)_{T\muC} =h$ (which can be
derived from \refeq{Gibbs--Duhem relation}).  Frumkin equation implies
that the slope of the line $\tensionC=\tensionC(\normalPC)$ at
constant $\muC$ is always positive and grows with the increasing $h$.
Moreover, the extrema of the functions $\normalPC(h)$ and
$\tensionC(h)$ at constant $\muC$ coincide, which causes the sharp
corners in the plot.

The line $\tensionC=\tensionC(\normalPC)$ in fig.\ \ref{chemical
potential loops} has numerous intersections, indicating a number of
possible combinations of coexisting phases.  In contrast, the results
in fig.\ \ref{volume fraction loops} have only one intersection.  For
a given system several equilibrium states may thus exist, but it may
be difficult to reach them in a stepwise-drainage process because of a
bottleneck at the second relaxation stage: the evolving coexisting
phases may encounter an instability before mechanical equilibrium
\refeq{equilibrium conditions for excess quantities} is reached.
After one or more of such transitions the film may break rather than
achieve a stable unconstrained-equilibrium configuration.  (Indeed,
film breakup was observed in most experiments with particle-stabilized
films; see, however, recent results reported in
\cite{Sethumadhavan-Nikolov-Wasan:2001}.)

Designing experiments that would clearly distinguish between different
evolution phases of the stepwise-drainage process is a challenge that
requires further numerical and theoretical studies.  Our approach will
be used in our forthcoming study to propose experimental protocols for
reaching full equilibrium states predicted by our analysis.

%
\acknowledgments{E.W.\ was supported by NASA grant NAG3-2704 and
J.B. was supported by NSF grants CTS-0201131 and CTS-S0348175.}


\end{document}